\documentclass[lettersize,journal]{IEEEtran}
\usepackage{amsmath,amsfonts}
\usepackage{algorithmic}
\usepackage{algorithm}
\usepackage{array}
\usepackage[caption=false,font=normalsize,labelfont=sf,textfont=sf]{subfig}
\usepackage{textcomp}
\usepackage{stfloats}
\usepackage{url}
\usepackage{verbatim}
\usepackage{graphicx}
\usepackage{cite}
\usepackage{enumitem}
\usepackage{multirow}
\usepackage{placeins}
\usepackage{mdframed}
\usepackage{booktabs}
\usepackage{xcolor}
\usepackage{tabularx}
\usepackage{hyperref}

\begin{document}

\title{From Hype to Collapse: Investigating Rug Pull Scams on Solana}

\author{Jiaxin Chen, Ziwei Li, Zigui Jiang,~\IEEEmembership{Member,~IEEE,}  Ruihong He, Yantong Zhou,\\Jiajing Wu,~\IEEEmembership{Senior Member,~IEEE,} Zibin Zheng,~\IEEEmembership{Fellow,~IEEE} 

\thanks{The research is supported by the National Natural Science Foundation of China (62332004, 62472457 and 62372485). \textit{(Corresponding author: Zigui Jiang.)}} 

\thanks{Jiaxin Chen, Ziwei Li, Zigui Jiang, Ruihong He, Yantong Zhou, Jiajing Wu and Zibin Zheng are with the School of Software Engineering, Sun Yat-sen University, Zhuhai 519082, China.}
}



\maketitle

\begin{abstract}
Solana has experienced rapid growth due to its high performance and low transaction costs, but the extremely low barrier to token issuance has also enabled widespread Rug Pulls. Unlike Ethereum-based Rug Pulls, which often rely on malicious smart-contract logic, Solana's unified SPL Token program shifts fraudulent execution toward on-chain behavioral manipulation. However, existing research has not systematically examined these Solana-specific Rug Pull patterns, and no public Solana Rug Pull dataset is available for empirical research. To bridge this gap, we present a large-scale measurement study of Rug Pulls on Solana. We manually verify 68 community-reported incidents and curate a benchmark of 117 confirmed Rug Pull tokens, from which we distill three representative on-chain behavioral patterns: Freeze Authority Abuse, Liquidity Withdrawal, and Pump-and-Dump. Guided by these patterns, we design a behavior-guided candidate identification and human-validation pipeline. We apply this pipeline to 100,063 tokens newly issued on Orca, Raydium, and Meteora during the first half of 2025, identifying 76,469 Rug Pull tokens. A random manual audit of 382 samples estimates a labeling false-positive rate of 0.26\%, supporting the reliability of the dataset. We release the resulting dataset and use it to characterize the Solana Rug Pull ecosystem. Our analysis shows that Rug Pulls on Solana exhibit extremely short lifecycles, strong price-driven dynamics, severe economic losses, and highly organized group behaviors. These findings provide new insights into the Solana Rug Pull landscape and support the development of effective on-chain defense mechanisms.
\end{abstract}


\section{Introduction}
Solana~\cite{Solana_whitepaper}, a high-performance blockchain platform launched in 2020, has emerged as a cornerstone of decentralized infrastructure by integrating Proof-of-History (PoH) with Proof-of-Stake (PoS). This architecture enables high throughput, low latency, and extremely low transaction costs. As of 2025, Solana has recorded over 1.05 billion active on-chain addresses and has emerged as the most popular blockchain ecosystem throughout 2024 and 2025~\cite{Solana_rank,Solana_Annual_Revenue}.

At the same time, the minimal cost of token issuance, the absence of listing barriers, and the lack of mandatory audits have fostered a ``high-frequency issuance, rapid-turnover'' ecosystem on Solana. While this open environment catalyzes innovation, it also provides a fertile ground for malicious actors. Consequently, Rug Pulls have proliferated, evolving into a unique and massive fraudulent ecosystem specific to Solana.
\looseness=-1

\textit{Rug Pull} is a quintessential cryptocurrency fraud wherein project developers attract investor capital through deceptive marketing or market manipulation, only to maliciously sell off tokens or abruptly withdraw liquidity. Such actions cause the project to collapse and the token price to plummet, leaving investors with substantial losses. Rug Pull incidents are rampant on Solana, leading to severe economic consequences. Kalal et al.~\cite{Economics_of_RugPUll} highlighted that Solana exhibits the highest density of Rug Pulls across major blockchains: approximately 55\% of newly issued tokens or liquidity pools eventually manifest Rug Pull behavior. Furthermore, a recent report~\cite{Solidus_labs} indicates that within the Raydium Decentralized Exchange (DEX) alone, up to 93\% of the 388,000 analyzed liquidity pools were suspected of Rug Pull activities, with the largest single incident resulting in a \$1.9 million loss.

Extensive research~\cite{CRPWarner,Stop_Pulling,Analysis_ETH_BNB,Do_Not_Rug,Pied-Piper,SoK,Uniswap_scam_token,TokenScout,RugPull_Scammers,NFT_RugPull} has analyzed the behavioral patterns of Rug Pulls on Ethereum and proposed automated detection methods based on smart contract, transaction sequences, or deployer behavior. However, these approaches are largely tied to the Ethereum Virtual Machine (EVM) execution environment. Owing to architectural differences between Solana and Ethereum, including the programming model of Solana, account state structures, transaction execution flows, and multi-instruction atomic transaction patterns, existing Ethereum-centric methods cannot be directly applied to Solana. To the best of our knowledge, although prior studies~\cite{Economics_of_RugPUll,RugPull_meme} have included Solana in cross-chain analyses of Rug Pulls, no systematic study has specifically focused on Rug Pull behaviors in the Solana ecosystem or thoroughly examined their platform-specific mechanisms and execution characteristics. This gap is not merely a matter of incomplete academic coverage: without a targeted empirical understanding of Rug Pulls on Solana, ecosystem stakeholders lack a reliable basis for calibrating risk assessment services, designing timely warnings, and developing effective mitigation strategies. Therefore, investigating Rug Pulls on Solana and constructing a Solana-specific empirical foundation for behavior-guided analysis are of clear research and practical significance.\looseness=-1

\noindent\textbf{Our Work.} To address this gap, we approach the problem from a dataset construction and ecosystem measurement perspective. We first manually collect and verify fraudulent incidents from 68 real-world community reports and curate a manually verified benchmark of 117 confirmed Rug Pull tokens, from which we distill three representative on-chain behavioral patterns: \textit{Freeze Authority Abuse}, \textit{Liquidity Withdrawal}, and \textit{Pump-and-Dump}. In this benchmark, every confirmed Rug Pull token can be unambiguously assigned to one of these patterns according to its primary on-chain execution and profit-taking mechanism, which defines the behavioral scope of our subsequent large-scale dataset construction. Guided by these behavioral patterns, we design a behavior-guided candidate identification and human-validation pipeline that collects, normalizes, and labels candidate Rug Pull tokens directly from raw on-chain transaction and state data. Applying this pipeline to 100,063 tokens newly issued on Orca, Raydium, and Meteora between January 1 and June 30, 2025, we construct the first large-scale Solana Rug Pull dataset, comprising 76,469 candidate Rug Pull tokens whose labeling reliability is further validated through random manual auditing. We then perform a systematic measurement on this dataset along five dimensions: token naming characteristics, lifecycle evolution, social media propagation, economic loss scale, and the organizational structure of fraudulent syndicates.

\noindent\textbf{Contributions.} The main contributions of this paper are summarized as follows:

\begin{itemize}[leftmargin=*, topsep=0pt]
\item We systematically characterize the on-chain mechanisms of Rug Pulls on Solana, contrast them with their Ethereum counterparts in Section~\ref{sec:cmp_ETH_SOL}, and trace the full lifecycle from token issuance to scam execution. From this analysis we distill three representative behavioral patterns: \textit{Freeze Authority Abuse}, \textit{Liquidity Withdrawal}, and \textit{Pump-and-Dump}, and curate a manually verified benchmark of 117 confirmed Rug Pull tokens.

\item We design a behavior-guided identification and human-validation pipeline that operates solely on raw on-chain transaction and state data, and apply it to 100{,}063 tokens newly listed on Orca, Raydium, and Meteora in the first half of 2025. The pipeline yields 76{,}469 in-the-wild Rug Pull tokens whose labels are validated by random manual auditing at a false-positive rate of 0.26\%. Together with the 117-token benchmark, the resulting corpus constitutes the first publicly released large-scale Solana Rug Pull dataset.
\item We measure the Solana Rug Pull ecosystem along five dimensions: token naming, off-chain propagation, lifecycle dynamics, economic impact, and syndicate organization. Rug Pull tokens systematically exploit trending topics and brand impersonation, rely on short-lived or low-credibility off-chain presence, and exhibit extremely short lifecycles tightly coupled to the SOL price. We further quantify at least \$151M in direct losses across 7{,}322 profiting addresses and identify 78 large-scale syndicates organized into two recurring on-chain topologies, namely \textit{Star} and \textit{Cluster}, evidencing a maturing fraud supply chain on Solana.

\end{itemize}

\noindent\textbf{Organization.} The remainder of this paper is organized as follows. Section~\ref{sec:background} introduces some preliminaries related to this study. Section~\ref{sec:anatomy} characterizes the complete attack workflow together with the representative token-level behavioral patterns. Section~\ref{sec:identify} describes the behavior-guided pipeline used to construct the large-scale Solana Rug Pull dataset..Section~\ref{sec:rqs} analyzes Rug Pulls on Solana in terms of deception features, economic impact, and scammer behaviors. Section~\ref{sec:mitigation} discusses potential mitigation strategies, and Section~\ref{sec:reltaed_work} reviews the related work. Section~\ref{sec:discussion} presents the limitations of our work and future research directions. Finally, the paper is concluded in Section~\ref{sec:conclusion}.
\section{Background}

\begin{table}[t]
\centering
\caption{Comparison of Rug Pulls between Ethereum and Solana.}
\label{tab:eth_solana_rugpull}
\hspace*{-10pt}
\scriptsize
\scalebox{1.0}{
\begin{tabular}{l l c c c}
\toprule
\textbf{Category} & \textbf{Property} & \textbf{Ethereum} & \textbf{Solana} \\
\midrule

\multirow{3}{*}{Token issuance}
 & Custom token contract deployment & $\checkmark$ & $\times$ \\
 & One-click token launch platforms & $\times$ & $\checkmark$ \\
 & Extremely low token deployment cost & $\times$ & $\checkmark$ \\

\midrule
\multirow{3}{*}{Fraud mechanism}
 & Malicious contract logic & $\checkmark$ & $\times$ \\
 & Hidden functions in contracts & $\checkmark$ & $\times$ \\
 & Behavior-driven Rug Pull & $\checkmark$ & $\checkmark$  \\

\midrule
\multirow{2}{*}{Detection support}
 & Code analysis as primary signal & $\checkmark$ & $\times$  \\
 & Dedicated academic detection tools & $\checkmark$ & $\times$ \\

\bottomrule
\end{tabular}
}
\end{table}

\label{sec:background}

\subsection{Web3 and DeFi}
Web3 represents a decentralized internet built upon blockchain technology, aiming to grant users full ownership of their identities, data, and assets. Built on this foundation, Decentralized Finance (DeFi) reconstructs traditional financial services through smart contracts, enabling intermediary-free lending, trading, and asset management. In the DeFi ecosystem, DEXs serve as the central hubs for asset liquidity. Unlike the order-book model of traditional finance, mainstream DEXs employ the Automated Market Maker mechanism. Under this paradigm, liquidity providers deposit token pairs into liquidity pools managed by code, where trading prices are dynamically adjusted via deterministic mathematical formulas. While this permissionless design significantly lowers the barrier to token issuance and allows anyone to create tokens and trading pairs, it also enables malicious actors to easily introduce low-quality or fraudulent assets into the public market.

\subsection{Solana}
Solana is a high-performance public blockchain platform designed to address the scalability issues within the blockchain trilemma. By introducing PoH as a core component of its consensus mechanism, Solana achieves massive throughput and sub-second confirmation times. As of 2025, Solana has evolved into one of the most active blockchain ecosystems in the world. Its high concurrency and low latency have made it a major platform for Meme token issuance and high-frequency trading scenarios~\cite{Solana_Meme}.

\subsection{Rug Pull}
A \textit{Rug Pull} is a quintessential fraudulent scheme in the cryptocurrency domain. It involves project orchestrators attracting investor capital through deceptive marketing or market manipulation, followed by malicious actions such as dumping tokens or abruptly withdrawing liquidity for profit. These actions cause the project to collapse and the token price to depreciate severely, resulting in significant financial losses for investors. Unlike hacking incidents that exploit code vulnerabilities, Rug Pulls typically occur within a functional code framework. Based on the execution method, these scams are categorized into Hard Rugs, which rely on malicious logic embedded in contracts, and Soft Rugs, which exploit market mechanisms without technical exploits~\cite{certikWhatSoft}.

\medskip

\label{sec:cmp_ETH_SOL}
\noindent\textbf{Solana vs. Ethereum.} Existing studies~\cite{CRPWarner,Stop_Pulling,Analysis_ETH_BNB,Do_Not_Rug,Pied-Piper,SoK,Uniswap_scam_token,TokenScout,RugPull_Scammers,NFT_RugPull} on Rug Pulls have primarily focused on EVM-compatible blockchains, particularly Ethereum. However, due to differences in underlying architecture and token mechanisms, Rug Pulls on Solana differ substantially from those on EVM-compatible platforms. To systematically illustrate these differences, we use Ethereum as a representative EVM-compatible platform and compare Rug Pulls on Solana and Ethereum along two key dimensions, as shown in Table~\ref{tab:eth_solana_rugpull}.

\textbf{(1) The cost and technical threshold for issuing tokens on Solana are significantly lower than on Ethereum.} Using official tools like the SPL-Token CLI or one-click Meme token launchpads such as Pump.fun, users can generate and deploy a new token within seconds by simply providing a token name and an image at a negligible cost of less than \$20. In contrast, Ethereum typically requires developers to write and deploy custom smart contracts. This process is susceptible to gas fee fluctuations, incurs higher costs, and demands specialized development expertise, with fewer mechanisms for automated, one-click issuance. Consequently, Solana outpaces Ethereum in terms of issuance efficiency, speed, and scale.

\begin{figure}[t]
  \centering
  \hspace*{-15pt}
  \includegraphics[width=0.53\textwidth]{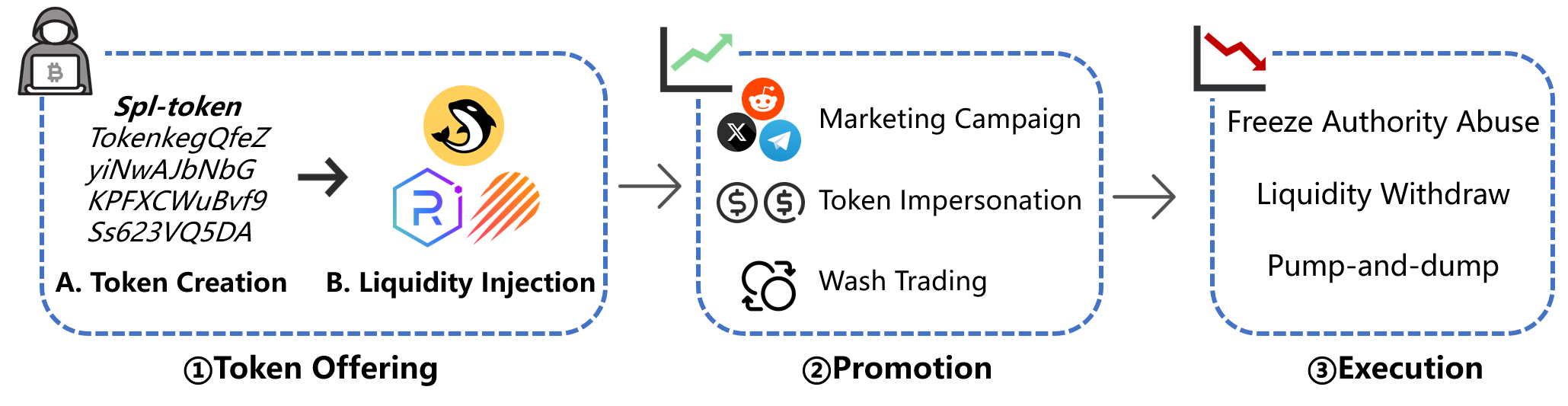}
  \caption{Workflow of Rug Pulls on Solana}
  \label{fig:Rug Pull process}
\end{figure}

\textbf{(2) Rug Pulls on Solana cannot be executed by modifying the contract code.} Many Rug Pulls on Ethereum exploit the programmability of smart contracts to embed malicious logic, such as reserving privileged functions that allow developers to restrict users from selling or modifying transaction tax rates. However, such attack vectors are infeasible on Solana. All Solana tokens are governed by the SPL-Token Program, meaning creators cannot modify the underlying execution logic of the token or embed custom backdoor functions as they can on Ethereum. Consequently, Rug Pull on Solana relies primarily on on-chain operational behaviors rather than the implementation of malicious program code.

Therefore, existing studies designed for detecting or analyzing Rug Pulls on EVM-compatible blockchains are not directly applicable to Solana due to fundamental differences in architecture and token mechanisms. This limitation highlights the necessity of dedicated research to systematically investigate the unique fraud mechanisms and on-chain behavioral patterns of Rug Pulls on Solana.

\section{Anatomy of Rug Pulls on Solana}
\label{sec:anatomy}

In this section, to understand the behavioral characteristics and token features of Rug Pulls on Solana, we manually collect and verify 68 fraud reports, identifying a total of 117 Rug Pull tokens. Based on these real-world cases, we summarize the overall operational workflow of Rug Pulls on Solana and further analyze the on-chain characteristics of these fraudulent tokens.
\looseness =-1

\subsection{Data Collection and Labeling}
We conduct a comprehensive search using the keywords \textit{Solana} and \textit{Rug Pull} across Chainabuse~\cite{chainabuse} reports as well as various analysis reports and incident retrospectives published by blockchain security teams. Chainabuse is a multi-chain fraud reporting platform co-maintained by multiple blockchain security institutions. Its reports typically include incident overviews, attacker addresses, relevant token information, and on-chain transaction links, providing direct and reliable leads for identifying Rug Pull tokens. Analysis reports from SlowMist~\cite{Slowmist} and CertiK~\cite{Certik} further supplement details such as attack workflows, fund flows, and risk patterns, improving the reliability of fraud labeling. Through this screening process, we extract and organize 117 malicious Rug Pull tokens and their associated information from 68 verified reports.

After completing data collection and screening, we recruit three researchers with over two years of experience in blockchain anti-fraud research to manually classify and label the identified Rug Pull tokens. We adopt an open card sorting~\cite{card_sorting} approach, constructing an analysis card for each fraudulent token that includes basic token information, the time of fraud occurrence, and the fraud type. Basic information including the token name, symbol, creation time, and holder count was exported directly via the Solscan API. The time of occurrence and fraud type are determined by researchers based on data from on-chain platforms, such as Solscan~\cite{solscan}, GMGN~\cite{gmgn}, Dextools~\cite{dextools}, Birdeye~\cite{birdeye}, and Dexscreener~\cite{dexscreener}. All analysis results are subject to independent assessments by the researchers and are finalized upon reaching consensus.

Finally, we construct a dataset that includes the time of fraud occurrence, token attributes, transaction behavior characteristics, and fraudulent tactics, providing a data foundation for subsequent Rug Pull token detection and analysis.

\subsection{The Workflow of Rug Pulls on Solana}
We analyzed 117 verified Rug Pull tokens and summarize a universal three-stage process followed by these scams: \uppercase\expandafter{\romannumeral 1}.~Token Offering, \uppercase\expandafter{\romannumeral 2}.~Promotion, and \uppercase\expandafter{\romannumeral 3}.~Execution, as illustrated in Fig.~\ref{fig:Rug Pull process}.

\bigskip
\subsubsection{\textbf{Token Offering}}
The first stage of a Rug Pull is the creation of a new token that can circulate and be traded in the market, which serves as the foundation of the scam. The Solana ecosystem is characterized by extremely low token creation costs, rapid deployment steps, and a lack of identity verification. Such characteristics contribute significantly to the high prevalence of Rug Pulls on the network. According to our categorization of real-world cases, we further subdivide this stage into two steps: Token Creation and Liquidity Injection.

\medskip

\noindent\textbf{Step 1: Token Creation.} In this sub-stage, fraudsters set the basic attributes of the token, including the name, logo, description, initial supply, related website or social media links. There are two primary avenues for token creation on Solana: the official Solana token tool SPL-Token and Pump.fun\footnote{Pump.fun is the fastest-growing Meme coin issuance platform on Solana known for its instant, low-cost, and no-code issuance mechanism. Despite massive daily token issuance volumes, most tokens on Pump.fun fail to establish liquidity pools and enter public markets.}.

A report~\cite{Solidus_labs} by Solidus Labs found that up to 98.6\% of tokens on Pump.fun become worthless pump-and-dump schemes shortly after launch. This implies that the vast majority of tokens on this platform not only have short lifecycles but also lack analytical value. Including these obviously fraudulent tokens in our research scope would yield a low return on computational investment and clutter the research sample with invalid noise, leading to results that fail to reflect Rug Pull behaviors with substantial impact. Therefore, we exclude Pump.fun-only tokens that never establish liquidity pools from our large-scale dataset construction. \textbf{For Pump.fun-originated tokens, we retain only those that are successfully issued and subsequently create liquidity pools.}

Token creation on Solana via the SPL-Token program is highly streamlined and cost-effective (approximately 0.21 SOL), allowing users to generate tokens and configure metadata with simple commands~\cite{pandatool}. Creators can retain sensitive administrative privileges, most notably the Freeze Authority. This permission enables the issuer to lock any account, preventing the holder from transferring or selling tokens. Scammers can exploit this mechanism by maliciously retaining the Freeze Authority to lock victim assets and execute Rug Pulls.
\looseness=-1

\medskip

\noindent\textbf{Step 2: Liquidity Injection.} Tokens created via the SPL-Token program do not possess native liquidity and therefore cannot be traded directly in on-chain markets. To make a token tradable, the issuer typically needs to create a corresponding trading pair on a DEX and inject initial capital into a liquidity pool, obtaining corresponding LP tokens. This process is a pivotal step in transforming a token into a tradable asset; once the liquidity pool is successfully created and activated, the token can be freely traded on the open market. LP tokens represent the proportional ownership of assets in a liquidity pool held by liquidity providers. Holders may redeem them at any time to withdraw the corresponding share of liquidity. As a result, the ownership and subsequent handling of LP tokens largely reflect the reliability of a token project. In Rug Pull schemes, project operators or affiliated addresses may deliberately retain LP tokens and later withdraw pool liquidity to realize rapid profits.

\bigskip
\subsubsection{\textbf{Promotion}}
After completing the token offering, the focus of the fraudsters shifts to rapidly inflating attention and trading volume to induce more victims to buy the token, thereby creating a liquidity base for the final cash-out phase. Case analysis indicates that fraudsters commonly employ three promotional tactics: Marketing Campaigns, Impersonation, and Wash Trading. These methods can be used individually or in combination to achieve more significant effects.

\medskip
\noindent\textbf{Tactic 1: Marketing Campaigns.} Fraudsters typically utilize mainstream online social networks (OSNs), such as X (formerly Twitter), Telegram, and Reddit to post content claiming ``the project is about to explode,'' creating an atmosphere of Fear Of Missing Out (FOMO) within the community to induce users to purchase the token. Some fraudsters purchase bot accounts to inflate likes and retweets, creating a facade of popularity. Moreover, some utilize community Key Opinion Leaders (KOLs) to promote tokens. These KOLs often have a strong influence on OSNs and can drive a significant number of Solana users to buy the fraudulent tokens. Fraudsters often secure KOL endorsements by stealing their accounts, paying for promotions, or even maintaining their own long-term KOL personas, thereby significantly enhancing the credibility and reach of the project.

\medskip
\noindent\textbf{Tactic 2: Impersonation.} Fraudsters may adopt impersonation strategies during the token creation stage, mimicking well-known tokens or elements to mislead victims into purchasing their tokens. Some tokens directly copy the logos of successful projects and appear in the transaction histories of victims through airdrop. Victims are likely to mistakenly purchase the token of the fraudster while intending to buy the legitimate one. Other fraudsters impersonate trending themes, faking team introductions and official websites to mislead victims into believing the token was issued by a popular team.

\medskip
\noindent\textbf{Tactic 3: Wash Trading.} After issuing the token, fraudsters may manipulate multiple accounts to frequently trade the token, artificially inflating trading volume to create an illusion of high demand. On Solana, transaction fees are extremely low, averaging 0.00025 USD~\cite{UnderstandSolanaTxs}, allowing fraudsters to execute wash trading at a very low cost. Wash trading across multiple accounts pushes up trading volume and reduces the proportion of the issued token in the liquidity pool, ultimately driving up the price and triggering FOMO in victims. Having multiple accounts within the fraud syndicate hold positions can also, to an extent, forge the illusion of a low top-holder concentration and high credibility.

Some sophisticated fraudsters choose Concentrated Liquidity Market Maker(CLMM) pools on Raydium. A feature of these pools is that liquidity providers can provide one-sided liquidity, adding only the project token without adding the quote token~\cite{raydium_CLMM}. The result of adding only one-sided liquidity is that any purchase of the issued token leads to a price increase, creating the illusion that the token price only goes up. This phenomenon easily attracts victims to buy the token.

\bigskip
\subsubsection{\textbf{Execution}}
The Execution stage is the final phase of a Solana Rug Pull. Its core characteristic is that the fraudster uses a series of techniques to rapidly cash out and abandon the project, causing heavy losses to users. In contrast to Ethereum, tokens on Solana are minted based on the official SPL-Token Program. Attackers cannot inject malicious logic into the program of the token itself, therefore Rug Pulls on Solana are generally not implemented through embedding malicious functions within token contracts. According to the characteristics of the Solana ecosystem, we categorize the final execution methods into three types: Freeze Authority Abuse, Liquidity Withdrawal, and Pump-and-Dump.

\medskip
\noindent\textbf{Type 1: Freeze Authority Abuse.} Token accounts on Solana have three distinct states: \textit{Uninitialized}, \textit{Initialized}, and \textit{Frozen}. Only the account holding the freeze authority can modify the state of different token accounts. A frozen account cannot perform any token-related operations, including transfers and swaps, resulting in a situation where the token can be purchased but cannot be sold.

Legitimate project teams typically renounce freeze authority during the token creation phase. However, Rug Pull operators can intentionally retain this authority. Once a victim swaps a quote token for the project token, the fraudster can invoke the \texttt{FreezeAccount} instruction to freeze the token account of victim, preventing any transactions. Meanwhile, accounts controlled internally by the fraudster remain free to sell tokens at any time, thereby directing the resulting profits to accounts controlled by the fraud syndicate.

\medskip
\noindent\textbf{Type 2: Liquidity Withdrawal.} This technique relies on the fraudster retaining LP tokens generated during the initial liquidity injection. In the early stages of token issuance, fraudsters establish a trading pool by injecting initial liquidity. This is followed by marketing activities and social media promotion to attract additional capital inflows from users, causing pool liquidity to surge and token price to rise. Once liquidity reaches a sufficient scale, the fraudster uses the retained LP tokens to redeem the assets originally injected.

\begin{table*}[t]
\caption{Statistical Comparison between Rug Pull and Legitimate Tokens}
\label{tab:cmp}
\centering
\small
\begin{tabular}{c c r r r r r r}
    \toprule
    
    \textbf{Statistic}   
    & \textbf{Type}  
    & \textbf{Lifespan(days)}
    & \textbf{Holders}
    & \textbf{Liq. Growth Ratio}
    & \textbf{DeFi Txs}
    & \textbf{Day-1 DeFi Ratio}
    & \textbf{Tx Rate (hr)} \\
    \midrule
    
    \multirow{2}{*}{Mean} 
    & Rug Pull   & 1.0339           & 83.18     & 42.20\%               & 4.17      & 95.69\%              & 0.21    \\
    & Legitimate      & 416.5262         & 215,952.89 & 657,910,914.63\%        & 184,802.68 & 3.40\%               & 3,463.85 \\
    \midrule
    
    \multirow{2}{*}{p25}  
    & Rug Pull   & 0.0024           & 5.00      & 6.15\%                & 0.00      & 100.00\%             & 0.04    \\
    & Legitimate      & 164.1428         & 11,505.00  & 1,764.71\%             & 26,812.25  & 0.00\%               & 121.90  \\
    \midrule

    \multirow{2}{*}{Median} 
    & Rug Pull   & 0.0116           & 9.00      & 7.60\%                & 2.00      & 100.00\%             & 0.08    \\
    & Legitimate      & 375.4549         & 41,026.00  & 4,636.00\%             & 92,362.50  & 0.00\%               & 299.11  \\
    \midrule

    \multirow{2}{*}{p75}  
    & Rug Pull   & 0.0576           & 24.25     & 21.37\%               & 2.00      & 100.00\%               & 0.22    \\
    & Legitimate      & 538.7596         & 100,787.00 & 28,047.62\%            & 234,893.25 & 0.06\%              & 1,436.01  \\      
    \bottomrule
\end{tabular}
\end{table*}

Since fraudsters usually hold a significant proportion of the LP tokens, the withdrawal operation rapidly depletes the liquidity pool. Without liquidity support, the token price collapses or drops to zero and market trading becomes dysfunctional shortly thereafter. Investors are unable to exit their positions and consequently incur substantial financial losses.

\medskip
\noindent\textbf{Type 3: Pump-and-Dump.} In these scams, fraudsters often employ more subtle market manipulation strategies. They may voluntarily renounce LP tokens or choose not to use the account freeze function to present the project as decentralized and fairly launched, thereby building user trust and attracting more participants. Once the project gains traction and the price rises, the fraudsters execute a concentrated sell-off using a large volume of pre-allocated tokens.

Some skilled fraudsters even disperse large amounts of tokens into multiple addresses to evade on-chain monitoring. When the token price reaches a target level, they liquidate their holdings at a high rate in exchange for mainstream assets. Due to the sudden influx of tokens into the market, the token price experiences a rapid decline and may approach zero, resulting in a severe devaluation of the holdings of ordinary investors.

\subsection{Token Behavior Characteristics}
To systematically analyze the differences between Rug Pull tokens and legitimate tokens on Solana and to inform the behavioral criteria used in large-scale candidate labeling, we construct a small-scale sample dataset of 182 tokens. The dataset consists of 117 Rug Pull tokens collected from previously identified fraud reports with clear evidence of malicious behavior, and 65 legitimate tokens verified by the isVerified mechanism~\cite{jup_Verif} of the Jupiter DEX Aggregator with an Organic Score~\cite{jup_score} of at least 80. Note that the selection criteria do not restrict the dataset to long-established assets or stablecoins. Specifically, among the legitimate tokens, five were created less than one month prior to the experiments, and twelve were created less than two months earlier, ensuring that the dataset also captures characteristics of newly issued yet legitimate tokens.

Following the preceding analysis of the Solana Rug Pull workflow, we classify each token in the manually verified benchmark according to its primary on-chain execution and profit-taking mechanism. The resulting taxonomy covers all 117 confirmed Rug Pull tokens in the benchmark: 2 tokens are classified as Freeze Authority Abuse Rug Pulls, 26 as Liquidity Withdrawal Rug Pulls, and 89 as Pump-and-Dump Rug Pulls.

After completing token labeling, we conduct a multi-dimensional comparative statistical analysis between fraudulent and legitimate tokens. The indicators cover four types of on-chain behavioral characteristics: lifecycle length, holder count, liquidity pool health, and the frequency and temporal distribution of DeFi behaviors, aiming to reveal the differences in their on-chain behavioral patterns. As detailed in Table~\ref{tab:cmp}, the statistical results reveal differences between Rug Pull and legitimate tokens, with Rug Pull tokens generally exhibiting a pattern characterized by extremely short lifecycles, rapid liquidity withdrawal, and a strong temporal concentration of transaction activity on the first day after deployment. Specifically, fraudulent tokens display a median lifecycle of merely 0.01 days and minimal holder counts, with the vast majority of their DeFi interactions occurring on the day of creation, accompanied by evident liquidity exhaustion. In contrast, legitimate tokens demonstrate sustained long-term activity and robust liquidity growth.

\textbf{Observations.} Rug Pull tokens on the Solana chain exhibit a pattern characterized by a short lifecycle and a strong concentration of activity on the first day in terms of lifecycle, holder growth, liquidity changes, and temporal distribution of behavior. This conclusion is consistent with our empirical observations of the Solana ecosystem and aligns with prior research from Cernera et al.~\cite{Analysis_ETH_BNB} on the behavioral characteristics of malicious tokens. This design prioritizes high-precision labeling of short-lifecycle Rug Pulls and intentionally treats slower variants as potential omissions.
\section{Dataset Construction Pipeline}
\label{sec:identify}

\begin{figure}[t]
  \centering
  \hspace*{-15pt}
  \includegraphics[width=0.53\textwidth]{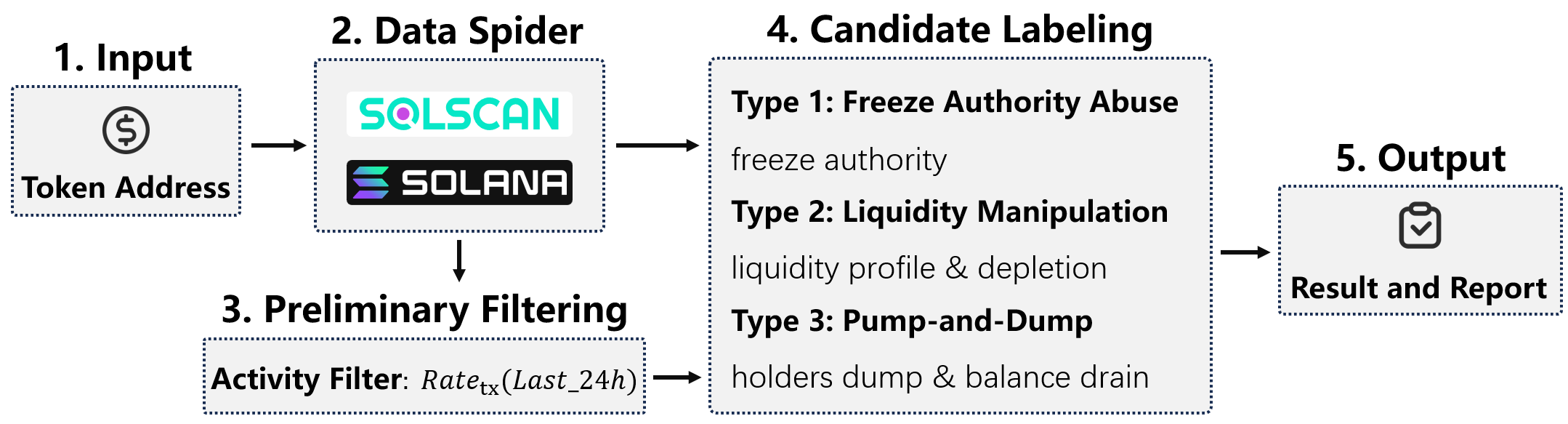}
  \caption{Overview of the behavior-guided dataset construction pipeline.}
  \label{fig:framework}
\end{figure}

Building on the three behavioral patterns characterized in Section~\ref{sec:anatomy}, we design a dataset construction pipeline  as shown in Fig.~\ref{fig:framework} that collects raw on-chain data, assigns candidate Rug Pull labels, audits these labels through random sampling, and releases a verified dataset. Throughout this process, we adopt a conservative labeling strategy that prioritizes label precision over coverage.

\subsection{Data Collection}
Token candidates are collected from three DEXes that dominate the Solana ecosystem: Orca~\cite{orca}, Raydium~\cite{raydium}, and Meteora~\cite{meteora}. These platforms jointly account for more than 80\% of Solana trading volume~\cite{Solana_dex_80,Solana_Dex_rank}, and thus provide broad ecosystem coverage for newly issued tokens. Between January 1 and June 30, 2025, after cleaning and deduplication, we obtain 100{,}063 newly issued tokens from these venues.

For each token, we retrieve the time-ordered transaction history, together with structured records of token creation, pool deployment, and liquidity changes, directly from the official Solana RPC and the Solscan API. The \texttt{getSignaturesForAddress} and \texttt{getTransaction} interfaces provide the raw transaction stream, while the \texttt{token meta}, \texttt{market info}, \texttt{defi activities}, and \texttt{token transfer} endpoints provide structured event records. We deliberately avoid aggregated or re-computed price feeds. A cross-check across GMGN~\cite{gmgn}, Birdeye~\cite{birdeye}, DexTools~\cite{dextools}, and DexScreener~\cite{dexscreener} shows that the reported price, percentage change, and FDV of the same token at the same timestamp often differ because of differences in trade-matching sources, estimation algorithms, pool-selection logic, and outlier handling~\cite{Stop_Pulling,Analysis_ETH_BNB,RugPull_meme}. By grounding every labeling decision in raw on-chain facts, the dataset can be reconstructed from the same primary sources.

\subsection{Behavior-Guided Labeling}
A coarse activity filter first removes tokens that are clearly still active. Specifically, a token passes this filter only if its average transaction rate over the trailing 24 hours is below 5 transactions per hour. This threshold is intentionally loose, and candidate inclusion remains stable when the threshold varies within the range of 0.5 to 5. The remaining tokens are then matched against the three behavioral patterns defined in Section~\ref{sec:anatomy}. \textit{Freeze Authority Abuse} requires that the creator retains the SPL-Token freeze authority and that at least one on-chain \texttt{FreezeAccount} instruction has been executed against a user account, thereby excluding dormant or formally renounced authorities. \textit{Liquidity Withdrawal} requires a creator-related address, namely the mint authority or the initial pool deployer, to execute a profitable liquidity removal that is followed by a collapse below the activity threshold, thereby distinguishing malicious exits from routine LP rebalancing. \textit{Pump-and-Dump} requires that, within a sustained post-launch window, both the holder count and the token balance in the pool decline monotonically, and that the overall holder reduction exceeds a threshold $\tau_{\text{down}}$. This criterion captures the characteristic ``cliff'' pattern without requiring an instantaneous collapse.\looseness=-1

\begin{figure}[t]
  \centering
  \includegraphics[width=0.5\textwidth]{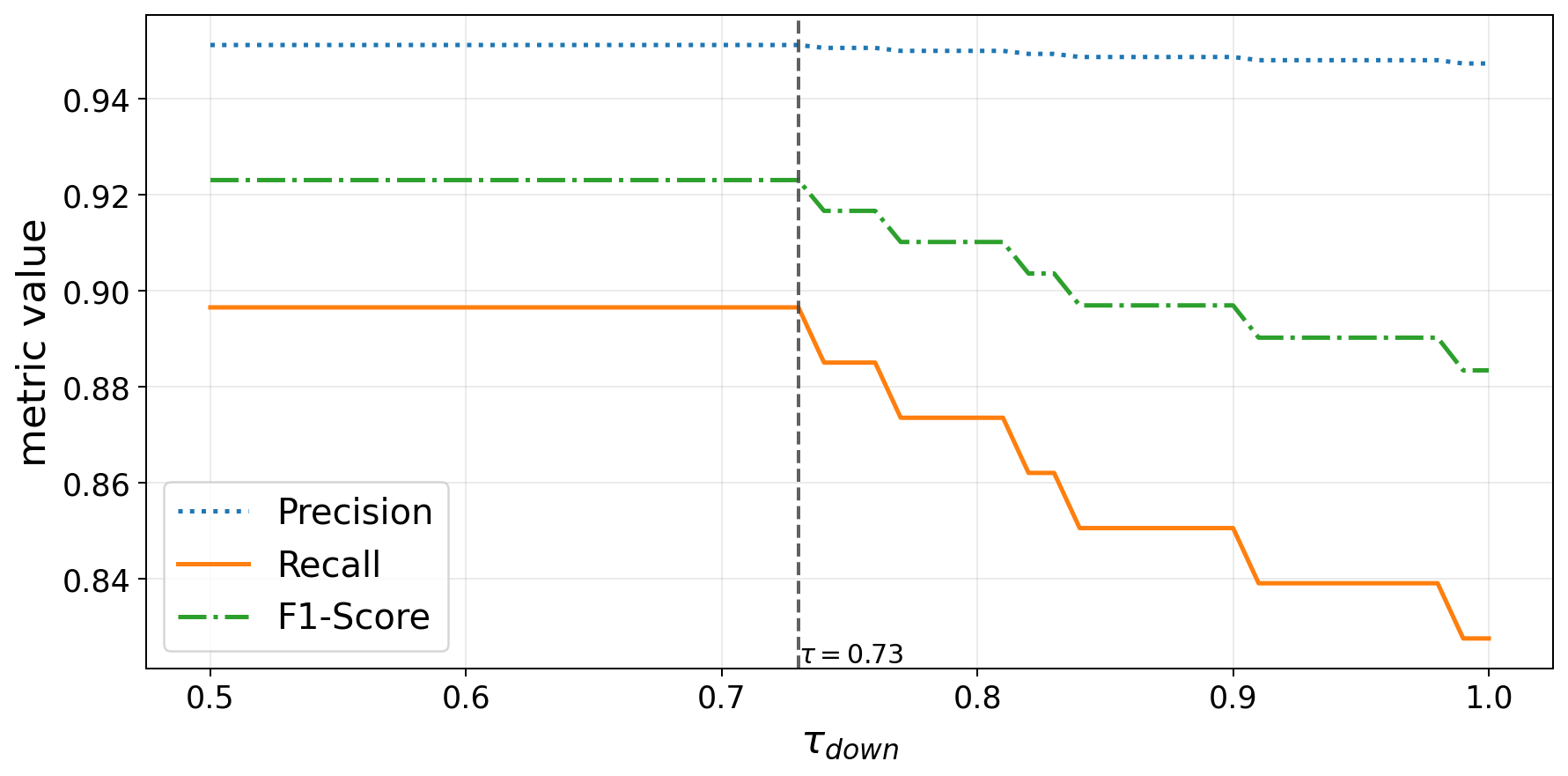}
  \caption{Labeling yield under different $\tau_{\text{down}}$ thresholds.}
  \label{fig:tau_cmp}
\end{figure}

Because false positives in labeling directly degrade dataset quality, $\tau_{\text{down}}$ is selected conservatively rather than tuned to optimize detection-style metrics. We sweep the threshold from 0.50 to 1.00 in increments of 0.01 (Fig.~\ref{fig:tau_cmp}). The candidate yield remains stable up to 0.73, but contracts sharply beyond this point as gradual sell-offs are filtered out. We therefore set $\tau_{\text{down}}=0.73$, which is the largest value that still admits gradual variants, and treat the resulting labels as a lower bound rather than an over-inclusive collection.

\subsection{Sampling-Based Quality Assessment}
Following standard practices in large-scale blockchain measurement studies~\cite{shaolost,SolPhishHunter,Uniswap_scam_token,RugPull_meme,NFT_RugPull}, we assess label quality through stratified random sampling rather than exhaustive enumeration. Specifically, we stratify the pipeline-labeled candidates by the three behavioral patterns used in our labeling process, namely \textit{Freeze Authority Abuse}, \textit{Liquidity Withdrawal}, and \textit{Pump-and-Dump}, and draw samples proportionally from each stratum. From the 76{,}469 pipeline-labeled candidates, we sample 382 tokens, accounting for approximately 0.5\% of all candidates. Under a binomial model, this sample size is sufficient to estimate the false-positive rate at the 95\% confidence level.

Three researchers with more than two years of experience in blockchain fraud analysis independently audit each sampled token. To make the manual assessment reproducible and reduce subjective judgment, we follow a structured validation protocol. For each token, the auditors review its token creation record, authority configuration, pool creation and liquidity operations, holder distribution changes, transaction sequence, price and trading dynamics, and post-event activity. A token is treated as a false positive only when its complete on-chain history does not satisfy the corresponding behavioral definition or when later evidence indicates that the suspicious behavior was not associated with a Rug Pull. After independent annotation, the auditors compare their results and resolve disagreements through discussion until a consensus label is reached.

Only one mislabeled token is identified, \textit{The First Revenue Backed Community Coin}\footnote{\texttt{zGCwX2FhMWakMfowT9D6kyLD2iU1cHe8Q94pt5cpump}}, which exhibited Rug Pull-like behavior at an early stage but was later revived through a community takeover. The observed false-positive rate is 0.26\%, with a 95\% upper bound of 1.45\%.

To further estimate omissions introduced by our conservative labeling strategy, we examine a complementary random sample of 100 tokens labeled as non-Rug Pull. Nine tokens are confirmed as missed cases, all of which unfold over three to seven days and therefore fall outside the 24-hour observation window. Among the correctly excluded tokens, some become malicious only after the labeling window closes, some contain too few transactions to form an identifiable pattern, and roughly 15\% remain in normal trading more than six months later. We report these numbers as dataset quality indicators under our conservative labeling strategy, rather than as detection performance.

\subsection{Released Dataset}
Across the 100{,}063 crawled tokens, the pipeline labels 76{,}469 tokens as Rug Pull candidates. These candidates consist of 60{,}402 \textit{Pump-and-Dump} tokens, 15{,}606 \textit{Liquidity Withdrawal} tokens, and 461 \textit{Freeze Authority Abuse} tokens. The publicly released Solana Rug Pull dataset comprises two complementary parts: \textit{Benchmark-117}, a manually verified ground-truth set of 117 tokens distilled from community reports in \S\ref{sec:anatomy}; and \textit{ITW-76469}, the 76{,}469 pipeline-labeled candidates audited above. Each entry contains the token address, behavioral category, and an on-chain event summary.
\section{Analysis of Rug Pulls on Solana}
\label{sec:rqs}
Using the constructed large-scale candidate dataset, this section analyzes the characteristics, operations, and impacts of Rug Pull activity on Solana. Focusing on the characteristics, operations, and impacts of Rug Pulls on Solana, we propose and explore the following three research questions:

\begin{itemize}[leftmargin=*, topsep=0pt]
\item \textbf{RQ1}: What are the deception characteristics of Rug Pull tokens on Solana in both on-chain and off-chain scenarios?
\item \textbf{RQ2}: What are the development patterns of Rug Pull on Solana, and what impact does it have on Solana?
\item \textbf{RQ3}: Is there a trend of syndicated, batched, or organized fraudulent activities in Solana Rug Pulls?
\end{itemize}

\begin{table*}[h]
    \centering
    \caption{Classification and Examples of Detected Anomalous Tokens}
    \label{tab:token_name_anomaly}
    \small
    \hspace*{-15pt}
    \begin{tabular}{cccll}
        \toprule
        \multirow{2}{*}{\textbf{Category}} & \multirow{2}{*}{\textbf{Count}} & \multicolumn{3}{c}{\textbf{Examples}} \\
        \cmidrule(lr){3-5}
        & & \textbf{Token Address} & \textbf{Token Name} & \textbf{Symbol} \\
        \midrule
        
        \multirow{6}{*}{\shortstack[l]{Inconsistent\\Metadata}} & 
        \multirow{6}{*}{3,203} & 
        \texttt{1239bythuxigVXPwfZkUk3Kvm8FS3KfxUsDUpzyPMAGA} & OfficialTrumpMelaniaBarron47MAGA & SOLANA \\
        & & \texttt{5gU9q7R4AhsStEUkqqootg76vyCzCu8qMiq2SjVwXGTR} & ElonMuskTrumpHarris69Inu & ETH \\
        & & \texttt{HLMcuxpo5VnY7nLV7CmbEv8KTyY7QgrbetYzs4Wve1uf} & 2) & USDC \\
        & & \texttt{4dHTQcvYcFvUqhnrYSFWReD14FhFUK5bko2LdKm8pump} & Official Useless Coin & ETH \\
        & & \texttt{13iyXpwj2NwQPjvmDoWdP4VhrPYouxHDdfgygFvepump} & I Think & SOAR \\\textit{}
        & & \texttt{9HqWnfE1vpmvbAcyLKA9GRgaaKeWArC44my85N3spump} & Bloom AI Assistant & VIDEO \\
        \midrule
        
        \multirow{5}{*}{\shortstack[l]{Look-alike\\Tokens}} & 
        \multirow{5}{*}{1,030} & 
        \texttt{82k69xiFtTBMkDdpV2u8ZcMKUnQ8ECS2NCCr7oAppump} & DeepSeek AI & DEEP \\
        & & \texttt{2A6GpST6j4kV13CHpL2N9V3vbKVEZXVwKZ1yTiSvpump} & USDT0 & USDT0 \\
        & & \texttt{BhEo2muprqjh43whgDz6yaBYA466dQfXHfVxVABrpump} & USDTea & USDT \\
        & & \texttt{BQDgRSCJ1V8Kx8fdvjqcwFa2cT7BUjQLRBpmh4rLQisD} & Official TIKTOK COIN & TIKTOK \\
        & & \texttt{FUHKqRi4nFYJ4Q2zgyQJrRGkx3KB99v2iPBkVgCzb3TT} & OFFICIAL TRUMP & TRUMP \\
        \bottomrule
    \end{tabular}
\end{table*}

\begin{figure}[t]
  \centering
  \includegraphics[width=0.5\textwidth]{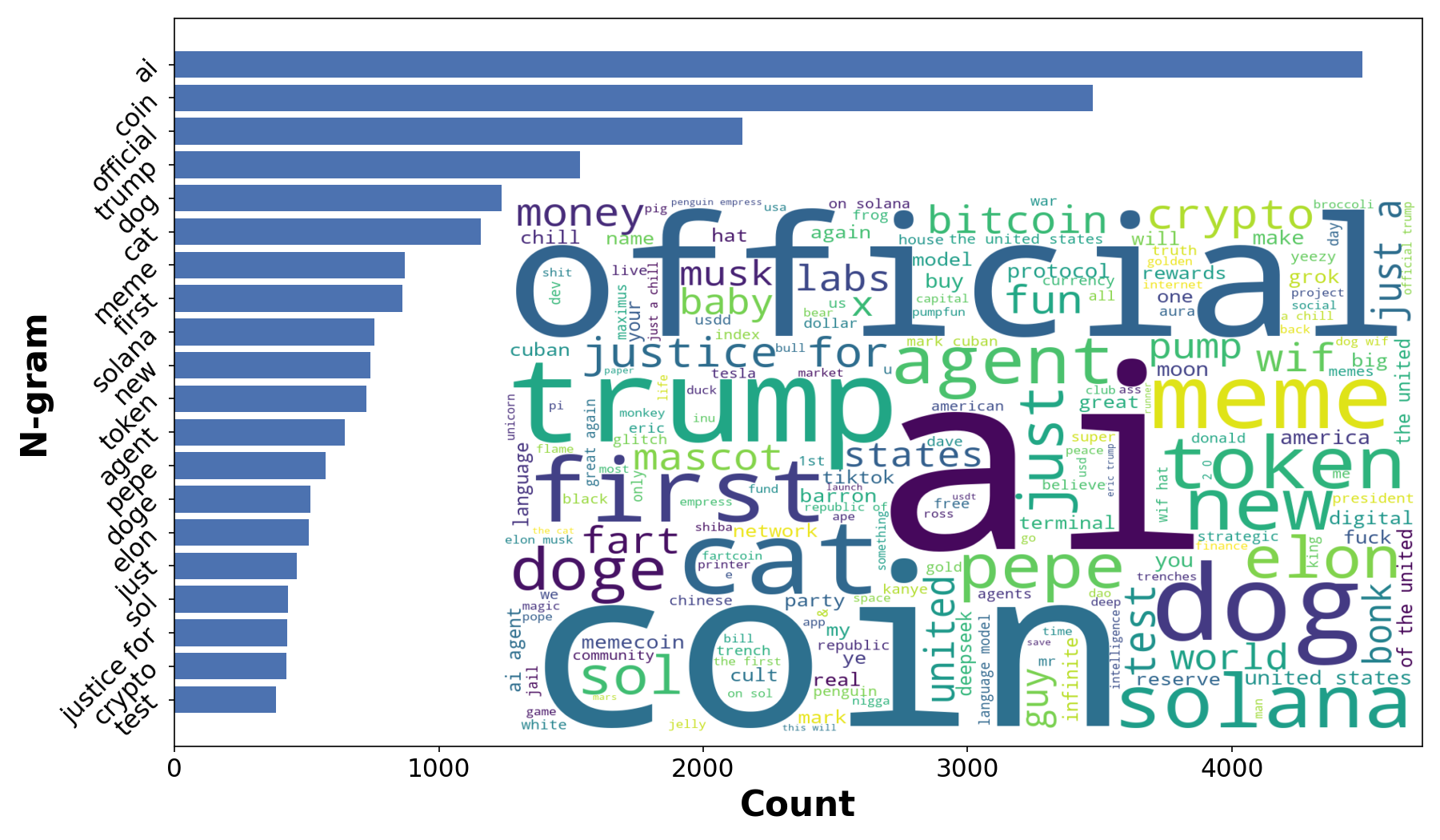}
  \caption{Most Frequent N-grams and Word Cloud of Rug Pull token Names on Solana}
  \label{fig:n-grams}
\end{figure}

\subsection{RQ1: Deception Feature Analysis}
\label{sec:rq1}
To characterize the deception features of Rug Pull projects on Solana at both the on-chain asset level and off-chain propagation level, we analyze from two complementary perspectives: the on-chain representation reflected by token names and metadata, and the off-chain propagation behavior of projects through official websites and social media.

\bigskip
\subsubsection{\textbf{On-chain Analysis}}
At the on-chain level, we first conduct name parsing and N-gram high-frequency statistics on the collected large-scale Rug Pull tokens. Overall, the naming of Rug Pull tokens exhibits a strong dependence on topical themes, accompanied by deceptive naming practices. As shown in Fig.~\ref{fig:n-grams}, from the perspective of global naming trends, AI and politically related themes dominate token names. Keywords such as ``AI,'' ``agent,'' and ``trump'' appear frequently, indicating that attackers intentionally leverage current hot topics to increase token exposure and trading attractiveness. Meanwhile, traditional pet-themed Memes, such as ``doge'' and ``pepe,'' maintain strong activity. The combination of classic Memes and emerging topics forms an overlapping naming structure, allowing Rug Pull tokens to maximize potential exposure and market attention.

Based on token naming analysis, we identify two highly representative malicious naming patterns with explicit deceptive intent, as summarized in Table~\ref{tab:token_name_anomaly}. The first pattern is characterized by deliberate inconsistencies between token names and symbols. In 3,203 cases, attackers deliberately assign token symbols that closely resemble mainstream assets or popular concepts (e.g., USDT, SOLANA, X, SORA), while the corresponding token names exhibit no clear semantic or functional association with these symbols. As many DEXs and wallets prioritize symbol display, this design can lead to visual misidentification during trading. The second pattern involves high-similarity impersonation through token naming. We identify 1,030 tokens whose names closely resemble well-known tokens or brands through direct imitation, suffix manipulation, or case-based confusion (e.g., USDTea, USDTf, TIKTOK). Such naming strategies exploit existing user trust and increase the likelihood of erroneous trading decisions.

\begin{figure}[t]
    \centering
    \includegraphics[width=0.5\textwidth]{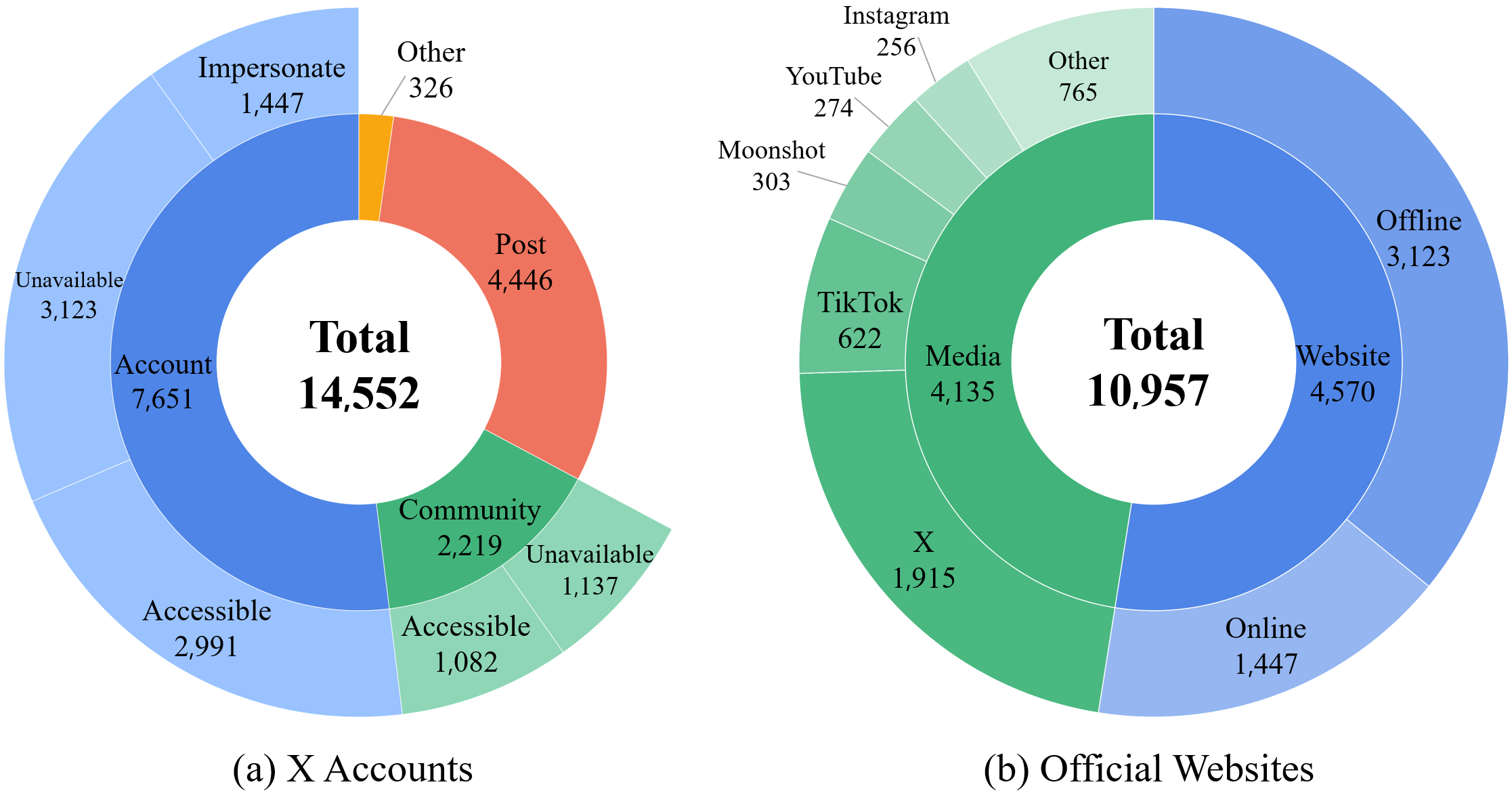}
    \caption{External Link Availability of Solana Tokens}
    \label{fig:offchain}
\end{figure}

\bigskip
\subsubsection{\textbf{Off-chain Analysis}}
At the off-chain propagation level, we utilize \texttt{Get Token Metadata} interface from Moralis~\cite{moralis} to collect social media information for 76,469 fraudulent tokens. Considering the functional differences between social media platforms, we focus our analysis on two highly representative dimensions: official websites and X accounts. These channels are selected because official websites are generally regarded as the most core and credible source of official information for cryptocurrency projects, while X has a significantly higher activity level and propagation speed in the crypto community compared to other platforms, serving as one of the primary promotional fronts for token issuance. Among 76,469 fraudulent tokens, only 14,552 (19.0\%) disclose an X account, and merely 10,957 (14.3\%) provide an official website link, indicating that most projects exhibit minimal or no off-chain presence.

For tokens that did provide off-chain links, both official websites and X accounts show strong signs of inconsistency and low credibility, as illustrated in Fig.~\ref{fig:offchain}(a) and Fig.~\ref{fig:offchain}(b). Among the 10,957 website links, only 29.3\% correspond to authentic and functional project websites, while over 70\% are either invalid or replaced by social media homepages and irrelevant sites. A similar pattern is observed on X. Among the 14,552 provided X links, over 60\% do not correspond to accounts with normal operational behaviors. These include links to celebrity or trending accounts (9.9\%), single-use promotional posts (30.6\%), and banned or deactivated accounts, with only 19.9\% exhibiting basic project-level maintenance and consistently low activity. Overall, these results indicate that off-chain identities in Solana Rug Pull projects are highly disposable and weakly bound to the token itself, which substantially limits the effectiveness of traditional social-signal-based risk assessment methods.\looseness=-1

\begin{mdframed}[linewidth=0.7pt, linecolor=black, skipabove=5pt, skipbelow=3pt, backgroundcolor=gray!6]
\textbf{Answer to RQ1}: Rug Pull tokens on Solana demonstrate systematic camouflage behaviors across both on-chain and off-chain dimensions. On-chain, their naming strategies implement visual deception through hot topics appropriation, name–symbol inconsistencies, and impersonation patterns. Off-chain, such projects generally lack authentic and sustained construction, with their social media activity characterized by a high degree of forgery, strong randomness, and short lifecycles. Overall, neither naming features nor off-chain signals alone enable reliable fraud identification, and it is imperative to conduct a comprehensive analysis by combining off-chain observations with on-chain behavioral analysis.
\end{mdframed}

\subsection{RQ2: Evolution and Impact Analysis}
\label{sec:rq2}
To characterize the survival status of Rug Pulls in the real Solana mainnet environment, we systematically analyze the developmental patterns of Rug Pull tokens and their impact on the ecosystem by combining temporal evolution features with economic loss quantification.

\begin{figure}[t]
  \centering
  \hspace*{-18pt}
  \includegraphics[width=0.53\textwidth]{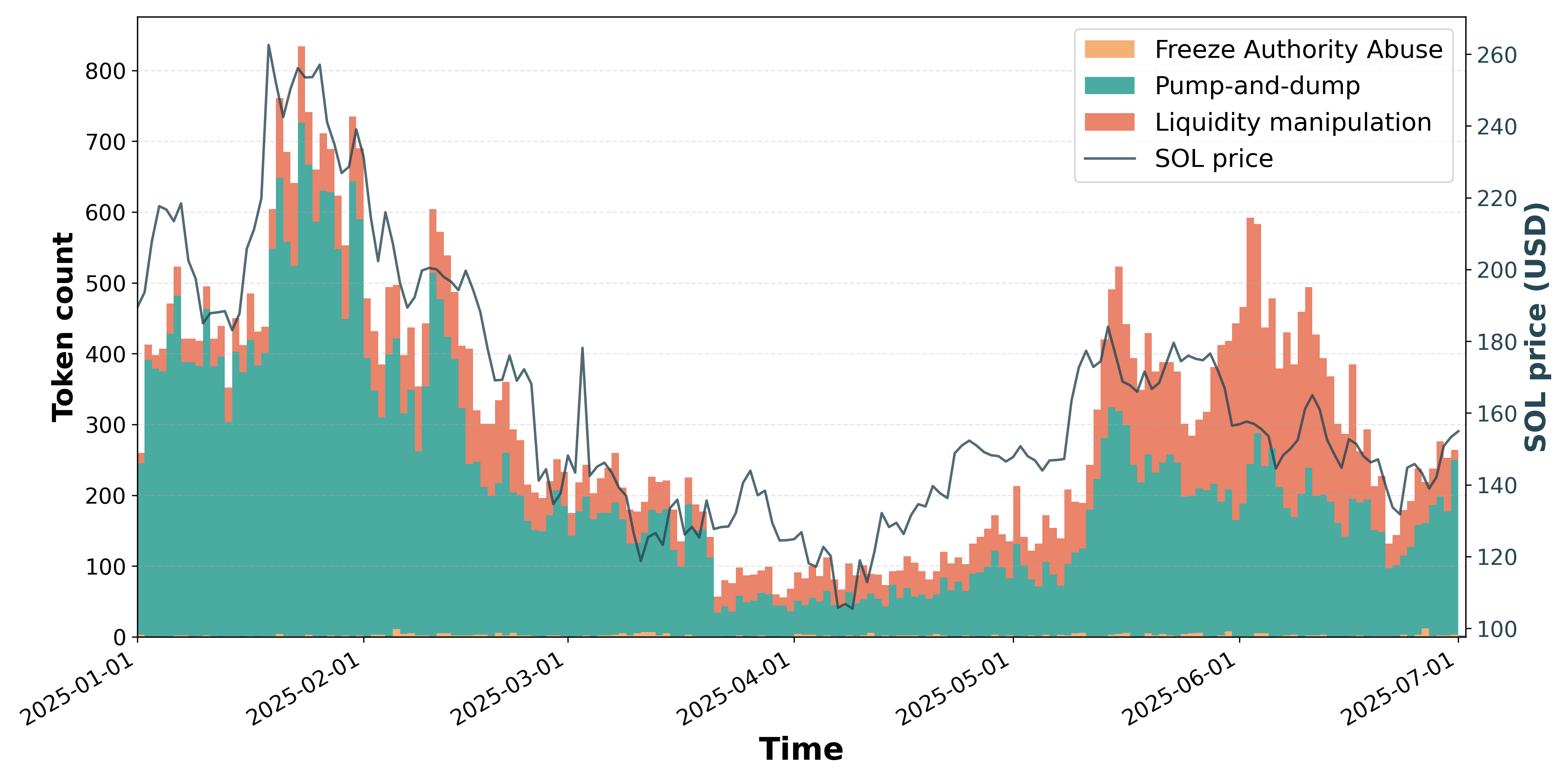}
  \caption{Evolution of Rug Pull Activities and SOL Market Price}
  \label{fig:evo}
\end{figure}

Regarding evolutionary characteristics, we define the lifecycle of a Rug Pull token as the time interval between its creation and the first confirmed fraud-related profit-taking transaction. As shown in Fig.~\ref{fig:evo}, combined with SOL token price data sourced from coingecko~\footnote{https://www.coingecko.com/en/coins/solana} for the first half of 2025, we find that the lifecycle of Rug Pull tokens on Solana is characterized by short durations and a high correlation with market price fluctuations. From the overall temporal trend, the volume of fraud exhibits a strong temporal correlation with the SOL spot price. When market prices rise or speculative sentiment is high, the number of executed fraud attacks tends to increase rapidly. This suggests that attackers strategically time the issuance of fraudulent tokens to periods of heightened market activity and speculative sentiment.

As shown in Table~\ref{tab:lifespan_stats}, Rug Pull tokens exhibit extremely short lifecycles. For Freeze Authority Abuses, Pump-and-Dump, and Liquidity Manipulation, the median lifecycle is less than one hour, and the 75th percentile of the lifecycle distribution generally does not exceed 5 hours. This suggests that even fraudulent tokens with relatively longer lifecycles complete profit extraction within a few hours, highlighting the high-frequency nature of Rug Pulls on Solana.

To quantify the economic impact of Rug Pulls, we adopt the on-chain profit tracing methodology proposed in the work of Xia et al.~\cite{Uniswap_scam_token}. Based on trackable on-chain profit paths, we quantify economic losses using liquidity changes of fraudulent tokens, measuring the actual amounts cashed out by attackers in two core assets: USDT/USDC and SOL. Specifically, we collect and analyze the DeFi activities of token and liquidity pool creators throughout the entire token lifecycle. By tracking balance changes resulting from their liquidity operations, we estimate the profits realized by attackers and the corresponding economic losses incurred by victims. Notably, the statistics in this study are limited to quantifiable and fully verifiable on-chain losses, defined as direct withdrawals of funds from victims by attackers. Therefore, the figures presented in this section should be regarded as conservative estimates of the economic impact of Rug Pulls.

As shown in Table~\ref{tab:financial_loss}, our in-the-wild detection results identify 7,322 profitable addresses associated with Rug Pull activity, resulting in at least \$151 million in quantifiable direct losses, highlighting the scale of the threat posed by Rug Pulls to the Solana on-chain economic ecosystem. The two profit asset paths exhibit different loss distributions. Stablecoin-based Rug Pulls display a strong head-concentration effect, whereas SOL-based Rug Pulls are characterized by numerous small-value, high-frequency scams with dispersed losses. This divergence may be partially explained by differences in token valuation, as SOL generally has a higher unit price than stablecoins, which constrains the scale of individual fraud events and encourages attackers to operate through numerous small-value transactions.

\begin{table}[t]
    \centering
    \small
    \caption{Statistical Distribution of Contract Lifespan by Attack Type}
    \label{tab:lifespan_stats}
    
    \hspace*{-10pt}
    \scriptsize
    \scalebox{1.1}{
    \begin{tabular}{lrrrr}
        \toprule
        \multirow{2}{*}{\textbf{Attack Type}} & \multicolumn{4}{c}{\textbf{Lifespan Statistics (Days)}} \\
        \cmidrule(l){2-5}
        & \textbf{Mean} & \boldmath{$P_{25}$} & \textbf{Median} & \boldmath{$P_{75}$} \\
        \midrule
        
        Rug Pull & 
        2.913668 & 0.008681 & 0.024618 & 0.064641 \\
        
        Freeze Authority Abuse & 
        14.558860 & 0.032749 & 0.483275 & 4.694080 \\
        
        Liquidity Manipulation & 
        9.873306 & 0.009039 & 0.038530 & 0.966377 \\
        
        Pump-and-Dump & 
        0.058967 & 0.008519 & 0.022083 & 0.048947 \\
        
        \bottomrule
    \end{tabular}
    }
\end{table}

\begin{table*}[t]
    \centering
    \caption{Financial Loss Analysis: Profitable Addresses and Volume Statistics}
    \small
    \label{tab:financial_loss}
    \begin{tabular}{l r rrrr r}
        \toprule
        \multirow{2}{*}[-0.5ex]{\textbf{Asset}} & 
        \multirow{2}{*}[-0.8ex]{\shortstack{\textbf{Profitable}\\\textbf{Addr.}}} & 
        \multicolumn{4}{c}{\textbf{Lost Amount (Native Units)}} & 
        \multirow{2}{*}[-0.8ex]{\shortstack{\textbf{Total Value}\\\textbf{(USD)}}} \\
        \cmidrule(lr){3-6}
        & & \textbf{Min} & \textbf{Max} & \textbf{Mean} & \textbf{Median} & \\
        \midrule

        USDT/USDC & 
        850 & 
        0.000001 & 10,565,400.13 & 10,805.06 & 38.08 & 
        17,774,329.19 \\
        
        SOL & 
        6,612 & 
        0.000001 & 4,576.74 & 59.96 & 5.99 & 
        133,315,108.61 \\
        
        \midrule
        \textbf{Total} & 
        \textbf{7,322} & 
        \multicolumn{4}{c}{N/A} & 
        \textbf{151,089,437.80} \\
        \bottomrule
    \end{tabular}
\end{table*}

\begin{mdframed}[linewidth=0.7pt, linecolor=black, skipabove=3pt, skipbelow=3pt, backgroundcolor=gray!6]
\textbf{Answer to RQ2}: Rug Pulls on Solana exhibit an extremely short lifecycle, with their issuance frequency highly synchronized with the SOL price. From an economic perspective, such scams have resulted in substantial financial losses, where attackers have cashed out at least \$151 million via stablecoin- and SOL-denominated transactions. Rug Pulls have thus become a critical risk threatening the stability and sustainable development of Solana.
\end{mdframed}

\subsection{RQ3: Scammers Analysis}
\label{sec:rq3}
To examine whether Rug Pulls on Solana demonstrate organized and collaborative behaviors, we adopt the definitions of fraudulent group roles proposed in the work of Huynh et al.~\cite{RugPull_Scammers}. We define token creators and liquidity providers who injected initial funds and extracted profits as \emph{core addresses associated with fraudulent tokens}, and cluster all addresses linked to these core entities. 

\begin{figure}[t]
    \centering
    \includegraphics[width=0.5\textwidth]{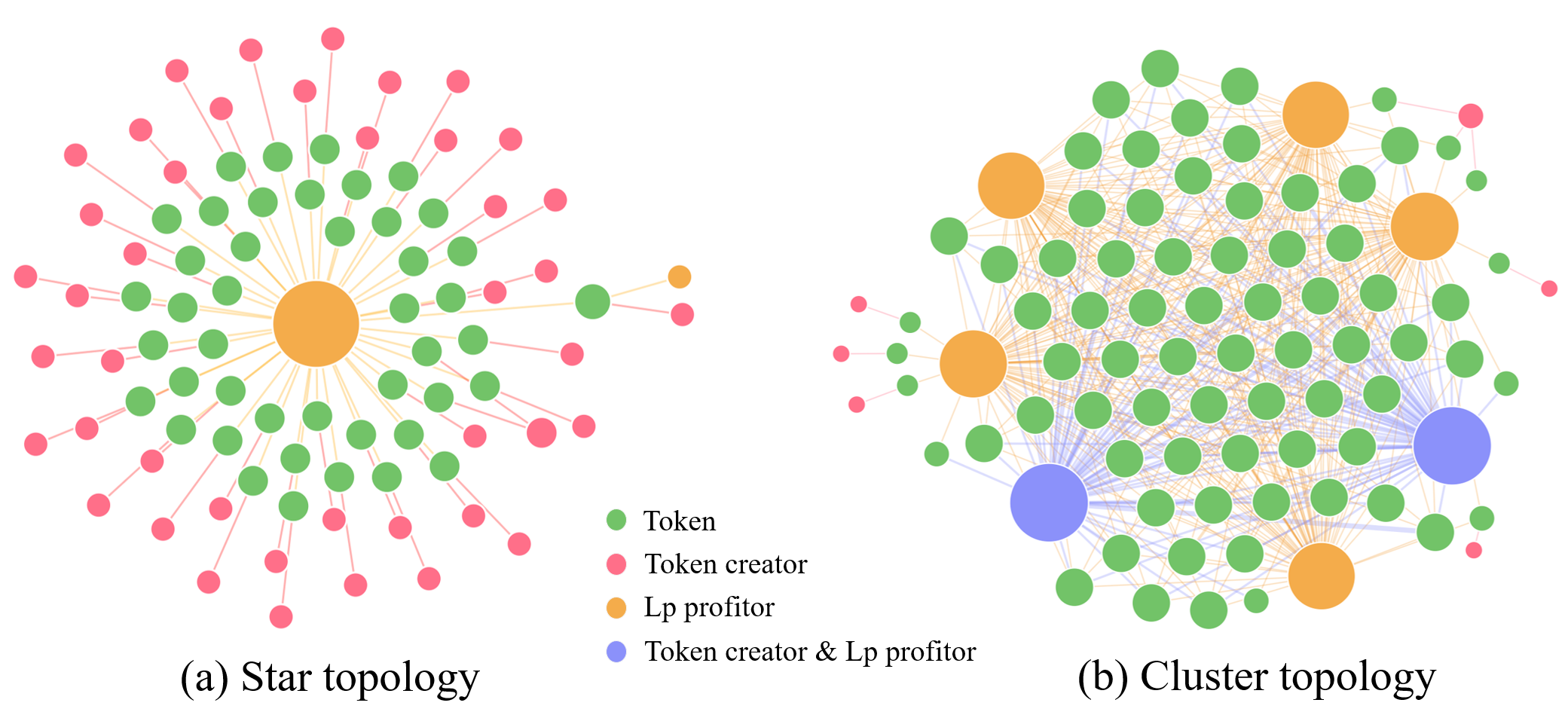}
    \caption{Network Topology Patterns of Rug Pull Schemes}
    \label{fig:scsmmers}
\end{figure}

We further restrict our analysis to groups in which the combined number of group members and associated tokens exceeds 50, treating them as candidates for organized fraud. This threshold is introduced to ensure that the analyzed samples exhibit sufficient scale and representativeness. Based on this criterion, we identify 78 large-scale groups, among which 16 correspond to single scammers repeatedly deploying multiple tokens from the same address, while the remaining 62 represent collaborative syndicates involving multiple participants. Overall, these groups exhibit clear organizational patterns and structured coordination behaviors on-chain. Within these syndicates, two typical on-chain organizational structures can be observed as illustrated in Fig.~\ref{fig:scsmmers}.

\begin{itemize}[leftmargin=*, topsep=0pt]
\item The first type is the \emph{Star Topology}, in which all illicit profits ultimately converge to a single central address, while numerous peripheral addresses including token creators, liquidity pool creators, and temporary operational addresses act as auxiliary nodes to enable high-speed, batch token deployment. In this structure, the core address typically conducts continuous, multi-cycle fraudulent activities through automated scripts, frequently deploying dozens or even hundreds of tokens within short time windows and executing highly repetitive Rug Pull operations.

\item The second type is the \emph{Cluster Topology}, which exhibits a more complex organizational structure, characterized by high-density and recurring cross-links among multiple key nodes, including token creators, liquidity injectors, and final profiting addresses. Different members do not merely engage in one-off collaborations but instead periodically co-participate in the deployment, manipulation, and profiting of multiple tokens. This model exhibits a clear division of labor; for example, some addresses primarily perform batch token creation, whereas others focus on initial liquidity injection, price manipulation, and profit aggregation.\looseness=-1
\end{itemize}

Table~\ref{tab:cluster_stats} presents descriptive statistics of the organizational structure and profitability of 78 large-scale fraud syndicates. Statistics indicate that these groups are highly organized in terms of both participant scale and token manipulation capacity. The number of members ranges from 1 to 169, with a median of 36 and a mean of 43.7, suggesting that most groups operate as stable collaborative networks rather than isolated individual actors. In terms of token manipulation scale, the data reveals a clear batch-based pattern: most syndicates do not profit from a small number of tokens but instead achieve sustained fraudulent output through large-scale, multi-batch token deployments. Overall, the statistics highlight an organized and systematic fraud model on Solana, characterized by large-scale token deployment and centralized profit extraction, which imposes sustained economic impact on the ecosystem.

\begin{mdframed}[linewidth=0.7pt, linecolor=black, skipabove=3pt, skipbelow=5pt, backgroundcolor=gray!6]
\textbf{Answer to RQ3}: Rug Pull on Solana has evolved into structured, organized on-chain operations. The Star Topology underpins a batch attack pattern controlled by a single core address, whereas the Cluster Topology represents a professionalized pattern featuring multi-account collaboration and clear division of labor. Drawing on the evidence from 78 large-scale syndicates, we confirm that Rug Pull on Solana has exhibited a distinct trend toward organized, batch-based, and systematic development.
\end{mdframed}

\begin{table}[t]
    \centering
    \caption{Descriptive Statistics of Gang Structure and Profitability}
    \label{tab:cluster_stats}
    \small
    \begin{tabular}{l rrrr}
        \toprule
        \multirow{2}{*}[-0.5ex]{\textbf{Metric}} & \multicolumn{4}{c}{\textbf{Statistics}} \\
        \cmidrule(l){2-5}
        & \textbf{Min} & \textbf{Max} & \textbf{Mean} & \textbf{Median} \\
        \midrule
        
        \multicolumn{5}{l}{\textit{\textbf{Network Structure (Counts)}}} \\
        \hspace{1em} Members & 1 & 169 & 43.70 & 36.00 \\
        \hspace{1em} Tokens Involved & 16 & 219 & 60.54 & 48.50 \\
        \hspace{1em} Creators & 1 & 169 & 32.20 & 25.00 \\
        \hspace{1em} Profitable Addrs & 1 & 53 & 14.45 & 12.00 \\
        
        \midrule
        
        \multicolumn{5}{l}{\textit{\textbf{Financial Profit (Value)}}} \\
        \hspace{1em} USDT/USDC & 0.47 & 794,205.13 & 36,669.34 & 138.11 \\
        \hspace{1em} SOL & 0.13 & 8,448.53 & 670.08 & 90.77 \\
        \hspace{1em} \textbf{Total (USD)} & \textbf{6.50} & \textbf{988,894.49} & \textbf{56,504.19} & \textbf{5,052.10} \\
        
        \bottomrule
    \end{tabular}
\end{table}
\section{Mitigations}
\label{sec:mitigation}
To address the prevalence of Rug Pulls and their economic impact within Solana, we derive several mitigation strategies from our dataset construction and measurement results.

\begin{itemize}[leftmargin=*, topsep=0pt]
\item \textbf{Data and service-level mitigations.} Risk assessment services should move beyond static token metadata and incorporate behavior-oriented signals from raw on-chain data. Our analysis shows that Rug Pulls on Solana are mainly executed through Freeze Authority Abuse, Liquidity Withdrawal, and Pump-and-Dump rather than malicious token contract logic. Service providers should therefore monitor token creation, pool deployment, liquidity operations, holder evolution, and suspicious fund flows as core risk signals.

\item \textbf{Interface and transaction-level mitigations.} DEXs, wallets, and aggregator front ends can reduce deception by improving token presentation and pre-transaction warnings. As shown in Section~\ref{sec:rq1}, attackers frequently exploit name--symbol inconsistencies, look-alike naming, and trending topics to impersonate mainstream assets. Front-end systems should highlight token addresses, display verified-address badges, and warn users when a token resembles a well-known asset but is not linked to an official verified address. Because Rug Pulls on Solana often have extremely short lifecycles, wallets and DEX interfaces should also provide lightweight warnings before swap confirmation when a token is newly created, has abnormal liquidity or holder dynamics, retains sensitive authorities, or is connected to suspicious addresses.

\item \textbf{Syndicate and user-level mitigations.} Defense mechanisms should account for the organized nature of Rug Pull activities. As shown in Section~\ref{sec:rq3}, many fraudulent tokens are associated with recurring address groups, centralized profit aggregation, and batch token deployment. Platforms can maintain relationship graphs connecting token creators, liquidity deployers, pool addresses, and profitable addresses, and raise stronger warnings when newly issued tokens match known high-risk groups or recurring deployment patterns. Users should also avoid relying only on token names, symbols, locked liquidity, or renounced authority, and should treat newly created social media accounts, inaccessible websites, and sudden hype around new tokens as risk signals.

\end{itemize}
\section{Related Work}
\label{sec:reltaed_work}
\subsection{Rug Pull}
Existing Rug Pull detection studies mainly fall into two categories: contract-level analysis and transaction-based analysis. 
Representative contract-level studies leverage static analysis techniques to uncover risky functions embedded in token contracts. Lin et al.~\cite{CRPWarner} applied bytecode analysis to extract suspicious manipulation functions. Zhou et al.~\cite{Stop_Pulling} summarized pump-and-dump risk patterns using Datalog-based contract analysis. Ma et al.~\cite{Pied-Piper} combined static analysis and fuzzing to identify common backdoors in ERC-20 contracts. Transaction-based approaches detect Rug Pulls using market indicators such as price, liquidity, and trading volume, through heuristic rules or machine learning models. Cernera et al.~\cite{Analysis_ETH_BNB} identified ``1-day Rug Pulls'' on Ethereum and Binance Smart Chain. Xia et al.~\cite{Uniswap_scam_token} extracted transaction features from Uniswap to detect fraudulent tokens. Mazorra et al.~\cite{Do_Not_Rug} combined contract and transaction features to train classifiers for scam detection. Wu et al.~\cite{TokenScout} modeled token activities as temporal graphs to enable early detection of emerging fraud.

Several studies further focus on specific ecosystems or attack perspectives. Huynh et al.~\cite{RugPull_Scammers} analyzed coordinated Rug Pull activities through cross-address correlation analysis. Li et al.~\cite{RugPull_meme} examined the trust ecosystem of Meme coins on Solana via sampled manual analysis. Huang et al.~\cite{NFT_RugPull} investigated exit-intent pump-and-dump patterns in Ethereum NFT markets.

\subsection{Solana Security}
Solana is a high-performance blockchain platform that has gained significant adoption in recent years. Its unique architecture provides high throughput while introducing new security and operational challenges. Regarding the Solana ecosystem, existing research has conducted analyses across multiple dimensions. Zheng et al.~\cite{Solana_why} conducted the first large-scale study of failed Solana transactions, identifying dominant error types and a 58.43\% failure rate among bots. Andreina et al.~\cite{Solana_dev} analyzed the security practices of Solana smart contracts from the perspective of a developer. In terms of underlying and automated analysis, Yuan et al.~\cite{Solana_eBPF} proposed the formal semantics of Solana eBPF. Cui et al.~\cite{VRust} and Smolka et al.~\cite{Solana_Fuzz} proposed vulnerability detection tools for Solana contracts from the perspectives of static analysis and binary fuzzing. Wu et al.~\cite{Solana_exp} systematically reviewed vulnerability types and security risks in Solana smart contracts. Furthermore, Li et al.~\cite{SolPhishHunter} presented the first systematic study of novel phishing attacks on the Solana.

\bigskip
Rather than merely migrating the detection of Rug Pulls from Ethereum to Solana, our work distinguishes itself by reformulating the underlying problem. Existing studies predominantly focus on EVM ecosystems, typically analyzing Rug Pulls through smart contract code, the behavior of deployers, or cross-chain statistical observations. Even the limited prior research involving Solana usually treats the platform as a single sample within broader cross-chain comparisons or focuses on general security issues, rather than specifically characterizing the execution mechanisms of Rug Pulls. In contrast, we conceptualize Rug Pulls on Solana as a market execution process organized around token lifecycles, DEX interactions, and the manipulatio

n of liquidity. Furthermore, we model this process from the perspective of account structures specific to Solana, program-driven interactions, and high-frequency trading environments. Building upon this perspective, we develop a reproducible evidence-driven dataset construction pipeline tailored to Solana and construct a large-scale dataset to support future research, thereby offering a research path different from the prevailing EVM-centered paradigm.
\section{Discussion}
\label{sec:discussion}
\noindent\textbf{Limitation.} Our pipeline is designed for high-precision dataset construction rather than real-time pre-attack warning. It assigns candidate Rug Pull labels based on observable on-chain transactions and state changes that have already occurred, which enables the resulting labels to be grounded in verifiable behavioral evidence. However, this design also makes the pipeline inherently retrospective. It is therefore better suited to constructing reliable in-the-wild labels and supporting ecosystem-level measurement than to warning users before a scam is executed. In addition, because we adopt a conservative labeling strategy and restrict the main observation window to 24 hours, the released labels should be interpreted as a high-precision lower bound rather than a complete census of all Rug Pulls on Solana. Slow-moving scams, extremely low-activity tokens, and cases whose malicious behavior emerges only after the observation window may be missed.

\noindent\textbf{Future work.} Future research may extend this work by exploring earlier risk assessment based on streaming transaction signals, finer-grained temporal features, and dynamic models of liquidity and holder evolution. Adaptive observation windows may also help capture slow-moving Rug Pull variants that unfold over multiple days while preserving label precision. In addition, learning-based models, temporal graph analysis, and cross-platform evidence from DEX interfaces or social channels can be combined with behavior-guided rules, provided that label verifiability and human validation are preserved.

\noindent\textbf{Ethical Considerations.} We emphasize that this study does not raise privacy concerns. First, all data used in our analysis are derived exclusively from publicly accessible transaction records on the Solana blockchain, which are available to any user. Second, the released dataset contains only anonymized on-chain transaction data and does not include any personally identifiable information. Based on these considerations, we believe that our study does not infringe upon individual privacy nor lead to legal or enforcement actions against any individuals.\looseness=-2

\section{Conclusion}
\label{sec:conclusion}
This paper presents a systematic empirical study of Rug Pulls on Solana from a dataset construction and measurement perspective. Starting from 68 community-reported incidents, we curate a manually verified benchmark of 117 confirmed Rug Pull tokens and characterize three representative behavioral patterns: Freeze Authority Abuse, Liquidity Withdrawal, and Pump-and-Dump. Guided by these observations, we design a behavior-guided candidate identification and human-validation pipeline based solely on raw on-chain transaction and state data. Applying this pipeline to 100{,}063 tokens newly issued on Orca, Raydium, and Meteora during the first half of 2025, we construct a large-scale Solana Rug Pull dataset containing 76{,}469 in-the-wild candidate tokens, with a manually audited false-positive rate of 0.26\%. Using this dataset, we further measure the Solana Rug Pull ecosystem across token naming, off-chain propagation, lifecycle dynamics, economic impact, and syndicate organization. Our results reveal systematic camouflage, extremely short lifecycles, price-driven dynamics, substantial financial losses, and organized group behaviors, providing a reproducible data foundation and empirical insights for behavior-oriented defenses in high-throughput, low-barrier blockchain environments.

\bibliographystyle{IEEEtran}
\bibliography{reference}

\begin{IEEEbiography} 
	[{\includegraphics[width=1in,height=1.25in,clip,keepaspectratio]{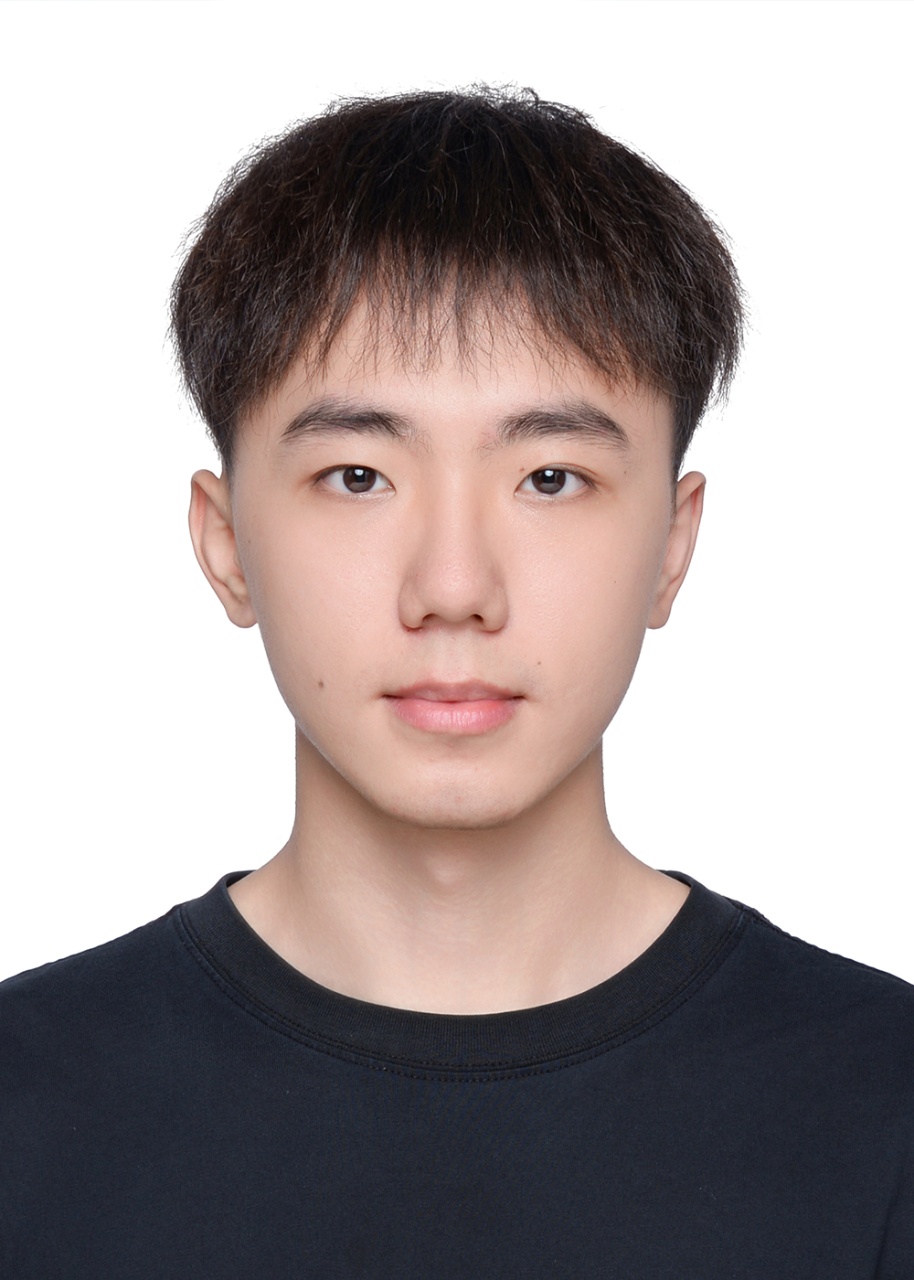}}]{Jiaxin Chen} received the B.Eng. degree in software engineering from Sun Yat-sen University, Zhuhai, China, in 2024. He is currently pursuing the M.Eng. degree with the School of Software Engineering, Sun Yat-sen University. His current research interests focus on blockchain security, including cryptocurrency fraud analysis, Rug Pull and phishing scam detection.
\end{IEEEbiography}

\begin{IEEEbiography} 
	[{\includegraphics[width=1in,height=1.25in,clip,keepaspectratio]{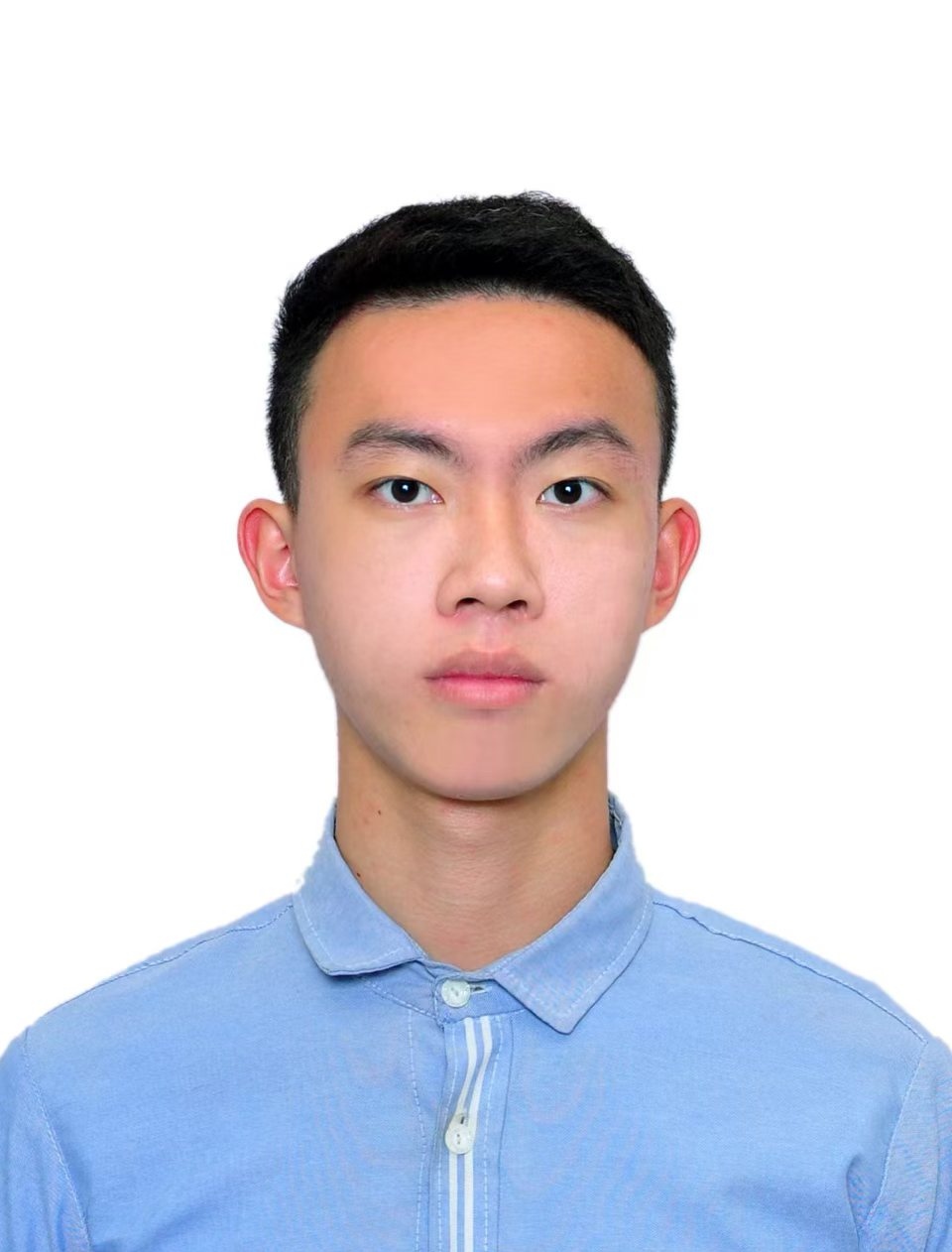}}]{Ziwei Li} received the B.Eng. degree in computer science and technology from Sun Yat-sen University, Guangzhou, China, in 2024. He is currently pursuing the Ph.D. degree with the School of Software Engineering, Sun Yat-sen University. His current research interests include blockchain, fraud detection, and network science.
\end{IEEEbiography}

\begin{IEEEbiography} 
	[{\includegraphics[width=1in,height=1.25in,clip,keepaspectratio]{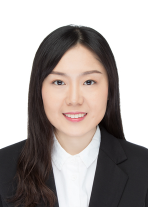}}]{Zigui Jiang} (Member, IEEE) received the Ph.D degree in computer science and technology from the Beijing University of Posts and Telecommunications, Beijing, China, in 2019. She is currently an Associate Professor with the School of Software Engineering, Sun Yat-sen University, China. Her research interests include blockchain, smart contracts, big data analysis, and service recommendation.
\end{IEEEbiography}

\begin{IEEEbiography} 
	[{\includegraphics[width=1in,height=1.25in,clip,keepaspectratio]{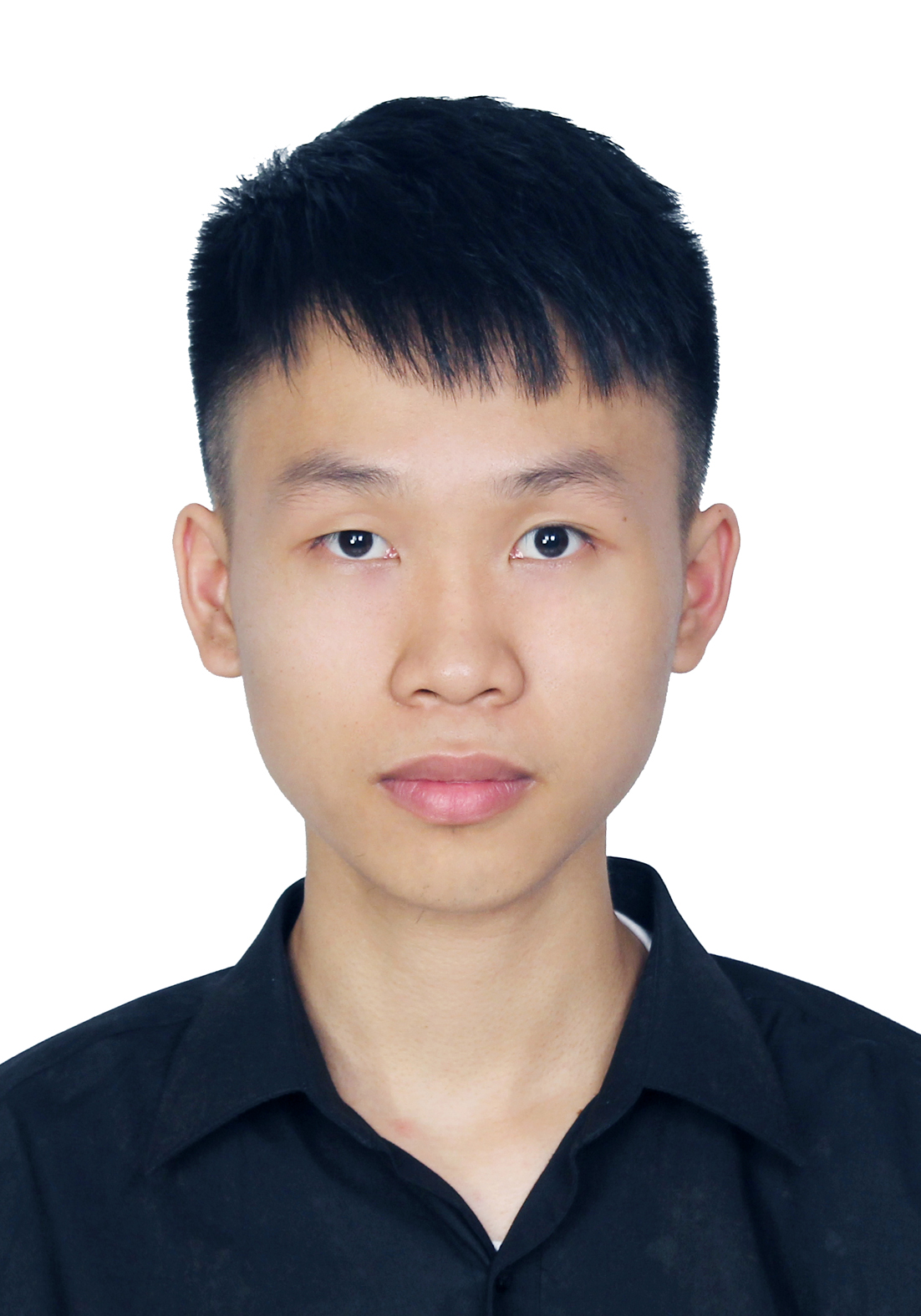}}]{
    Ruihong He} is currently pursuing the B.Eng. degree in software engineering with the School of Software Engineering, Sun Yat-sen University, Zhuhai, China. His current research interests include blockchain security, decentralized finance fraud analysis, and rug pull detection.
\end{IEEEbiography}

\begin{IEEEbiography}
    [{\includegraphics[width=1in,height=1.25in,clip,keepaspectratio]{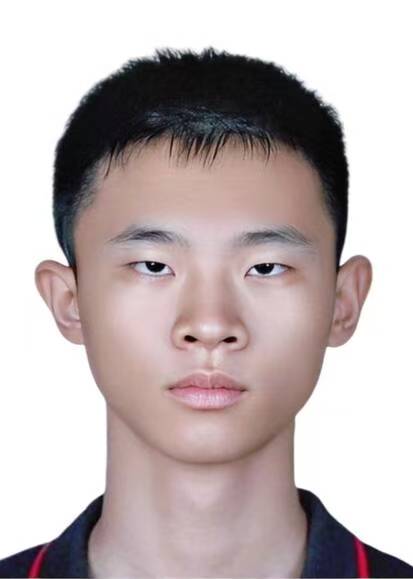}}]{Yantong Zhou} is currently pursuing the B.Eng. degree in software engineering with the School of Software Engineering, Sun Yat-sen University, Zhuhai, China. His current research interests focus on blockchain security, malicious activity detection, and stablecoin de-pegging risk analysis.
\end{IEEEbiography}

\begin{IEEEbiography} 
	[{\includegraphics[width=1in,height=1.25in,clip,keepaspectratio]{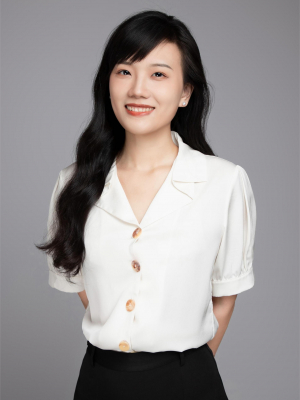}}]{Jiajing Wu} (Senior Member, IEEE) received the B.Eng. degree in communication engineering from Beijing Jiaotong University, Beijing, China, in 2010, and the Ph.D. degree from Hong Kong Polytechnic University, Hong Kong, in 2014. She was awarded the Hong Kong Ph.D. Fellowship Scheme during her Ph.D. study in Hong Kong (2010--2014). 
    She is currently a Professor and the Deputy Dean with the School of Software Engineering, Sun Yat-sen University, Zhuhai, China. Her research focus includes blockchain, graph mining, network science. She serves as an Associate Editor for {\sc IEEE Transactions on Circuits and Systems II: Express Briefs.}
\end{IEEEbiography}

\begin{IEEEbiography}
	[{\includegraphics[width=1in,height=1.25in,clip,keepaspectratio]{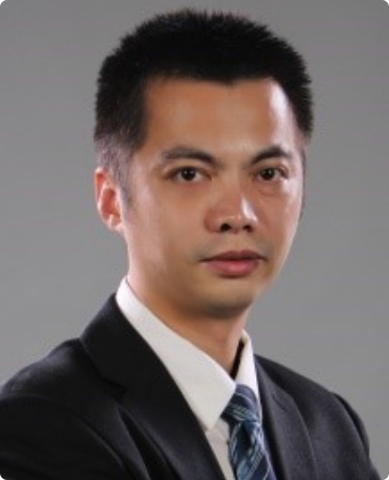}}]{Zibin Zheng}(Fellow, IEEE) is currently a professor and the dean with the School of Software Engineering, Sun Yat-sen University, Zhuhai, China. He authored or coauthored more than 200 international journal and conference papers, including one ESI hot paper and ten ESI highly cited papers. According to Google Scholar, his papers have more than 28,000 citations. His research interests include blockchain, software engineering, and services computing. He was the BlockSys’19 and CollaborateCom16 General Co-Chair, SC2’19, ICIOT18 and IoV14 PC Co-Chair. He is a Fellow of the IET. He was the recipient of several awards, including the Top 50 Influential Papers in Blockchain of 2018, the ACM SIGSOFT Distinguished Paper Award at ICSE2010, the Best Student Paper Award at ICWS2010. He is a fellow of the IET.
\end{IEEEbiography}

\end{document}